\documentclass[prl,aps,twocolumn,showpacs,preprintnumbers,amsmath,amssymb,superscriptaddress]{revtex4}

\usepackage{graphicx}	% Include figure files

\begin{document}

\title{ Connection between supernova shocks, flavor transformation, and the neutrino signal}

% the following should give superscripted affiliations
\newcommand{\UCSD}{Department of Physics, University of California, San Diego,
La Jolla, California 92093-0319}
\newcommand{\LANL}{Los Alamos National Laboratory, Los Alamos, NM 87545}

% author list
\author{Richard C.~Schirato}\affiliation{\UCSD}\affiliation{\LANL} 
\author{George M.~Fuller}\affiliation{\UCSD}

\date{\today}

\begin{abstract}
We show that the supernova neutrino signal in terrestrial detectors could give characteristic signatures as the shock propagates through the regions where matter-enhanced neutrino flavor transformation takes place.  This effect could allow existing and proposed supernova neutrino detectors to provide new insights into the problem of shock re-generation and propagation.  Alternatively, such observations could allow unique insights into the neutrino mass and mixing matrix.
\end{abstract}

%Physics and Astronomy Classification Scheme

\maketitle

In this letter we point out a link between the expected neutrino signal from a core collapse supernova event on the one hand, and neutrino mass/mixing physics and supernova shock re-heating and propagation physics on the other.  The crux of the supernova problem is understanding how the initial core bounce shock (which stalls at time {\it post core bounce} $t_{\rm pb} \sim 0.1\,{\rm s}$) becomes re-energized by neutrino heating (at $t_{\rm pb} < 1\,{\rm s}$) \cite{SNmodels}.  Conventional observations (optical, radio, and x-ray) of supernova events and remnants give little direct information on the re-heating history and early propagation of the shock.  Here we demonstrate that the neutrino signal may show characteristic time structure in energy/luminosity stemming from shock propagation across the (resonance) regions where matter-enhanced neutrino flavor transformation takes place.  In principle, observations of such feaures could provide insight into the shock strength and location as functions of time. The experimentally-determined mass and mixing properties for the active neutrinos suggest that these flavor transformation-generated features coincidentally occur when the neutrino luminosity is high.  Detection of such shock-related time structure in the supernova neutrino signal, in turn, could yield information on neutrino mass and mixing parameters complimentary to that derived from experiment. 

Solar and atmospheric neutrino experiments have revealed that neutrinos have vacuum masses and that they may mix with large (perhaps near maximal) mixing angles \cite{solar,atmos}. The LSND experiment \cite{lsnd} can be interpreted as evidence for additional vacuum neutrino mixing in the $\bar\nu_\mu\rightleftharpoons\bar\nu_e$ channel with a characteristic mass-squared difference $\delta m^2 \sim 1\,{\rm eV}^2$. This result will be checked in the near future by the MiniBooNE experiment \cite{boone}. If this signal holds up, then in combination with the solar and atmospheric neutrino anomalies we are forced to adopt new physics: either the introduction of a light singlet (\lq\lq sterile\rq\rq ) neutrino which mixes with the active species; or CPT-violating schemes in which neutrinos and antineutrinos have different masses and mixing properties \cite{cpt}. If the former, then there are two mass/mixing arrangements which have been advanced as giving simultaneous solutions for all the neutrino anomalies: the \lq\lq 3~+~1\rq\rq\ scheme wherein the fourth (heaviest) neutrino mass state could be mostly singlet in character; and the \lq\lq two-doublet\rq\rq\ scheme in which a lower mass doublet is involved in the solution of the solar neutrino problem and the higher mass doublet is invoked for the atmospheric anomaly and the LSND signal.  In any case, our neutrino flavor transformation/shock propagation features in the supernova signal will arise for the pure active (3 neutrino) scenario as well as for those involving sterile neutrinos.

Matter enhancement of neutrino flavor conversion in the supernova \cite{SN} will be engendered by these neutrino mass schemes via the Mikheyev-Smirnov-Wolfenstein (MSW) mechanism \cite{msw,q&f95}.  MSW resonances in the region above the hot proto-neutron star could alter the energy spectra and fluxes of the expected standard supernova neutrino signal. Neutrino flavor conversion effects on the supernova neutrino signal have been considered \cite{oldSNsig,FHM} previously, but without consideration of shock effects.

In fact, the shock represents a major perturbation on the pre-supernova density and electron fraction ($Y_e$) profiles. Furthermore, the shock is expected to be moving through the region above the neutron star where neutrinos with typical energies ($E_\nu \sim 10\,{\rm MeV}$) will be resonant at the times ($t_{\rm pb} \approx 0.1\,{\rm s}$ to $\approx 15\,{\rm s}$) when neutrino fluxes are large enough to be potentially detectable. Neutrino flavor conversion efficiency at resonance depends on the density/electron fraction gradient.  Since the shock front is very steep, neutrino flavor conversion can be \lq\lq shut off\rq\rq\ as the shock passes through an MSW resonance region.  Subsequently, as the shock moves beyond the resonance region, flavor conversion could be reestablished, albeit at a smaller radius, likely corresponding to a higher density gradient.  The net result could be that, for example, $\nu_{\mu,\tau}\leftrightharpoons \nu_{\rm e}$ flavor conversion is \lq\lq interrupted\rq\rq\ for a time of order the shock (front and rarefaction zone) propagation across resonance regions.  This can give dips or steps in neutrino energy/count rate in terrestrial detectors. 

Core collapse supernova simulations including numerical treatments of hydrodynamics and neutrino transport are not always in agreement. However, there are essential features of shock physics that are common to these models: a rapid (instantaneous) density jump, followed by a rarefaction region where the density can drop well below that at the leading edge of the shock. We have drawn on results from several calculations \cite{wilson2} to produce a generic time-dependent fit to the density profile immediately above the neutron star and through the region of the shock. The density profiles at several times are shown in Fig.~1. We have employed a $\rho \sim 1/r^{2.4}$ profile \cite{wilson2} for both the progenitor star and the \lq\lq neutrino-driven wind\rq\rq\ which forms above the neutron star at $t_{\rm pb} > 2\,{\rm s}$. (The results are similar if we use a $\rho \sim 1/r^3$ profile as suggested by the progenitor model of Ref.\ \cite{heger}.) We have steepened the shock front in our density profiles relative to that suggested by numerical calculations. Shock fronts in hydrodynamics calculations may be softened by numerical techniques, leading to unphysically large neutrino flavor conversion probabilities there. 

\begin{figure}
\includegraphics[width=3.4in]{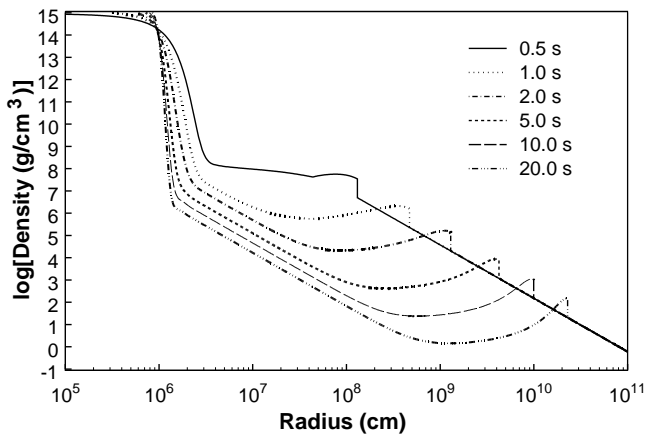}
\caption {Adopted density profile for progenitor and shock wave as a function of radius at various times {\it post-bounce} $t_{\rm pb}$.}
\end{figure}

We take the neutrino energy distribution functions at the neutrino sphere to be Fermi-Dirac black bodies with small chemical potentials, $f(E_\nu)= \left( 1/T_{\nu_\alpha}^3 F_2(\eta_{\nu_\alpha})) E_\nu^2/(\exp{(E_\nu/T_{\nu_\alpha}-\eta_{\nu_\alpha})} + 1\right)$, for neutrino species $\nu_\alpha$ with $\alpha =e,\mu,\tau$ (assuming no singlet neutrino flux from the neutron star surface). Here average neutrino energies are given in terms of neutrino sphere temperatures by $\langle E_{\nu_\alpha}\rangle = T_{\nu_\alpha} F_3(\eta_{\nu_\alpha})/F_2(\eta_{\nu_\alpha})$ with $F_k$ the standard relativistic Fermi integral of order $k$. For illustrative purposes, we have adopted neutrino degeneracy parameters $\eta_{\nu_\alpha} = 3$ for all neutrino species and have taken $ \langle E_{\nu_{\mu ,\tau}}\rangle =\langle E_{\bar\nu_{\mu ,\tau}}\rangle = 26.9\,{\rm MeV}$, $\langle E_{\bar\nu_e}\rangle = 16.0\,{\rm MeV}$, and $\langle E_{\nu_e}\rangle = 10.0\,{\rm MeV}$ for all times \cite{qian93}.  In fact, neutrino degeneracy parameters and average energies will evolve with time. At early times it may be that all neutrino species have similar energy spectra, but at late times when the core has substantially de-leptonized, they may be different as above. 

The energy differential fluxes for each neutrino species are given in terms of the luminosities $L_{\nu_\alpha}$, neutrino sphere radius $R_\nu$, and radius from the center of the neutron star $r$ by $d\phi(E_\nu) \approx (L_{\nu_\alpha}/2\pi R_\nu^2)[(1-\sqrt{1-R_\nu^2/r^2})/{\langle E_{\nu_\alpha}\rangle}]f(E_{\nu}) dE_\nu$. We take the liberated gravitational binding energy as $3\times 10^{53}\ \rm {ergs}$; and neutrino luminosities to be the same for all active species with an exponential decay time constant of $\tau = 3\,{\rm s}$.

Neutrino flavor amplitude evolution here is calculated as in Ref.s\ \cite{qian93,q&f95}, in an assumed succession of two-by two transformations:
\begin{equation}
\label{ampevol}
 i \frac {d}{dt} \left [\begin{array}{c}
a_{\alpha} \\ a_{\beta}
\end{array}\right ]   =   \left [ \begin{array}{cc}
A + B - \Delta \cos 2\theta & B_{\rm off} + \Delta \sin 2\theta \\
B_{\rm off} + \Delta \sin 2\theta & \Delta \cos 2\theta - A - B \end{array} \right ] \left [\begin{array}{c} a_{\alpha} \\ a_{\beta} \end{array}\right ]
\end{equation}
where $a_\alpha$ and $a_\beta$ are the amplitudes for the neutrino to be flavor $\nu_\alpha$ or $\nu_\beta$, respectively; $\Delta = \delta m^{2}/2E_\nu$ with $\delta m^{2}$ the appropriate neutrino mass-squared difference. The neutrino potentials which drive flavor evolution are: $A=\pm \sqrt{2}G_{\rm F} n_{\rm b}Y_e$, where $n_{\rm b}$ is the baryon number density; $B$ and $B_{\rm off}$ are the flavor diagonal and off-diagonal neutrino-neutrino forward scattering potentials, respectively. In the calculations presented here we neglect the neutrino-neutrino forward scattering contributions to the potentials. While justified where the neutrino fluxes are low; at early times, near the neutron star, or where the entropy is high, these terms (as well as feedback on $Y_e$) can be important in determining MSW flavor transformation probabilities \cite{FMBF}.  These effects can alter our time-dependent spectral features in some cases \cite{schirato}.

We discuss the supernova neutrino signal resulting from four distinct neutrino mass and mixing schemes: (1) 3-active neutrinos only, with a \lq\lq normal\rq\rq\ mass hierarchy; (2) \lq\lq 3~+~1\rq\rq\ with $m_{\nu_4} \gg m_{\nu_3}$; (3) the two doublet scheme; and (4) a CPT-violating scheme with the neutrinos as in (1) but with an inverted mass hierarchy in the anti-neutrino sector producing additional resonances. We find that shock effects are important in setting the neutrino signal in {\it all} of these neutrino mass schemes.

For the 3-active neutrino scheme and typical density profiles across and behind the shock front, there will arise a sequence of MSW resonances (each in the $\nu_{\mu,\tau}\leftrightharpoons\nu_{e}$ channel): up to three resonances associated with the atmospheric mass-squared difference scale (taken to be $\delta m^2 = 3 \times {10}^{-3}\,{\rm eV}^2$); and up to three associated with the solar $\delta m^2$-scale.  (For all cases we take the Large Mixing Angle solar solution and adopt $\delta m^2 = 3.7 \times {10}^{-5}\, {\rm eV}^2$ and $\sin ^{2}2\theta_{\odot} = 0.8$.)  The resonance density is $(\rho Y_{e})_{res} \approx 6.55\times{10}^{6}\,{\rm g}\,{\rm cm}^{-3} \delta m^{2}\cos2\theta /E_\nu$ where $\delta m^2$ is in  ${\rm eV}^2$ and $E_\nu$ is in $\rm MeV$.  For demonstration purposes, we take $\theta_{13}=0.07$.  It should be noted that several features in the neutrino signals are sensitive to the choice of $\theta_{13}$, possibly allowing a probe of this parameter.  

The \lq\lq 3 + 1\rq\rq\ scheme has an additional set of higher-density resonances associated with the LSND $\delta m^2$-scale in the $\nu_e\rightleftharpoons\nu_s$ channel.  In this case, $Y_e$ is replaced by $3(Y_{e}-1/3)/2$.  (For illustrative purposes we adopt $\delta m^2 = 1.0\,{\rm eV}^{2}$ and $\sin ^{2}2\theta_{\rm LSND} = 3.5\times 10^{-3}$ for all LSND-scale two-by-two resonances, where $\theta_{\rm LSND}$ should be understood to be an effective two-by-two vacuum mixing derived from the appropriate 4-neutrino unitary transformation.) For the two-doublet scheme we follow Ref.s\ \cite{bf_cfq} and exploit the (near) maximal $\nu_\mu\rightleftharpoons\nu_\tau$ mixing to treat flavor evolution above the neutron star as two sequential two-by-two level crossings: $\nu_{\mu,\tau}\rightleftharpoons\nu_s$ followed at larger radius (lower density) by $\nu_e\rightleftharpoons\nu_{\mu,\tau}$. At even larger radius there can be a matter-enhanced transition at the solar $\delta m^2$ scale.  Given the shock profiles as in Fig.~1, each level crossing could give up to three resonance regions.  We take an inverted mass hierarchy for the antineutrino sector of case (4).  The LSND $\delta m^2$-scale is associated with the  $\bar\nu_e \rightleftharpoons\bar\nu_{\mu , \tau}$ resonance in this case.

We calculate the evolution of the neutrino flavor states through our adopted density profile (including the time dependent position of the shock) by numerically integrating Eq.\ \ref{ampevol} using an adaptive Runge-Kutta procedure or the semi-classical Landau-Zener approach \cite{LZ} (with subsequent projection onto vacuum flavor eigenstates) where applicable. As the shock moves out and overtakes the MSW resonance regions appropriate for each neutrino mass scheme, the neutrino energy spectra for $\nu_{\mu,\tau}$ and $\nu_e$ will evolve as a result of shock modification of the density profile. In particular, the rarefaction zone behind the shock may introduce additional level crossings (resonances) that would be absent in the smooth, monotonic density run of the progenitor star. 

Fig.~2 shows charged current (CC) event rates for $\nu_e$ and $\bar \nu_e$ and total neutral current (NC) rates on deuterium \cite{deutcross} as functions of time from a 10 kpc distant supernova for the four mass/mixing schemes discussed above.  These rates are calculated for 0.77 kT of $\rm {D}_2 \rm O$, similar to the fiducial mass in the SNO detector.  Ratios of CC to NC rates are plotted in Fig.~3.  The variation in time (and associated model-dependent uncertainties) of the neutrino emission rates from the core are thereby suppressed.  Statistical fluctuations in the signals are neglected, and we take $Y_e = 0.5$ in this simplified example.  All four mass/mixing schemes show deviations in time from a simple exponential for the event rates, or from a flat line for the ratios.  The two doublet scheme shows a dip in the $\nu_e$ CC event rate centered at approximately 1 s.  When integrated over the full width, this dip represents a 33\% reduction out of an estimated 58 events over the same time period without this effect.  The CPT violating scheme exhibits a 37\% reduction out of $\sim$46 events in the $\bar \nu_e$ CC rate for a dip also centered at about 1 s.  The 3-active, \lq\lq 3 + 1,\rq\rq\ and CPT violating cases show similar dips in the $\nu_e$ CC rate in the range $5\ {\rm s} \lesssim t_{\rm pb} \lesssim 11\ {\rm s}$ representing about a 20\% reduction out of $\sim$60 events.   

\begin{figure}
\includegraphics[width=3.5in]{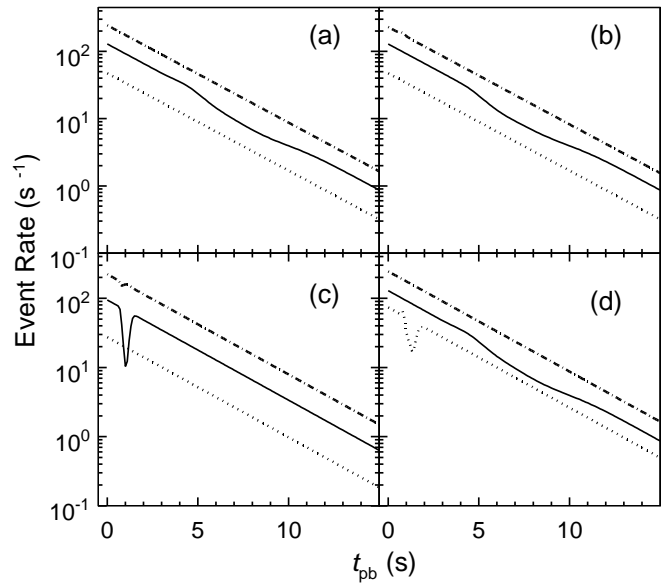}
\caption {Event rates for $\nu_e$ CC (solid), $\bar \nu_e$ CC (dotted), and NC (dot-dash) in SNO as functions of time for: (a) 3-active, (b) \lq\lq 3 + 1,\rq\rq\ (c) two doublet, and (d) CPT violating schemes.}
\end{figure}

\begin{figure}
\includegraphics[width=3.5in]{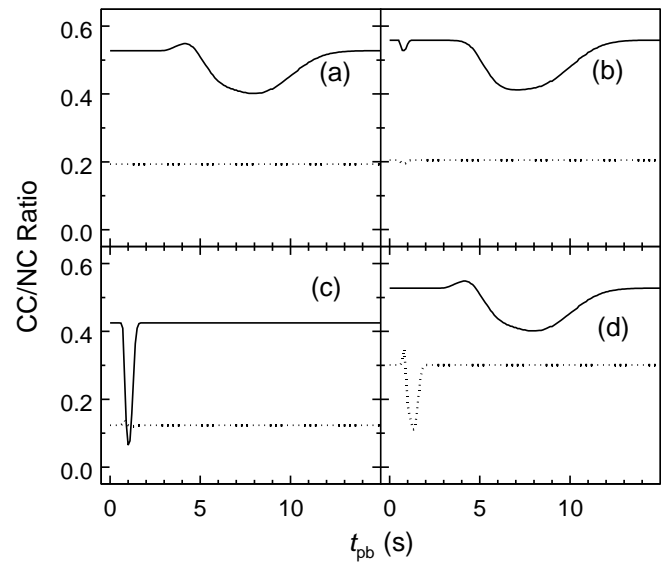}
\caption {Ratios of $\nu_e$ (solid) and $\bar \nu_e$ (dotted) CC to NC event rates for SNO.  Neutrino mixing schemes are (a) 3-active, (b) \lq\lq 3 + 1,\rq\rq\ (c) two doublet, and (d) CPT violating.}
\end{figure}

While the timing of the structures in the CC signals are dependent upon the assumed shock model, we expect that such structures will be a generic feature of the neutrino burst from a core-collapse supernova. NC rates show time structure for the cases which include a sterile neutrino, e.g. two doublet and \lq\lq 3~+~1\rq\rq, though at low amplitudes for the parameters discussed here.  The 3-active and CPT-violating cases would show no shock-related time dependence in the NC signal since the NC interactions are \lq\lq flavor blind.\rq\rq\ Improved signal statistics could be obtained by considering detectors with substantially increased target mass.  The use of light/heavy water and -CH$_2$ as supernova neutrino detection media are discussed in Ref.s\ \cite{SKSNOdetect,Cdetect}.  The proposed use of high atomic number materials such as lead \cite{FHM,leaddetect} could significantly improve the detectability of the structures in the $\nu_e$ CC signals.

For the results shown in Figs.~2 and 3, we have assumed all resonances are adiabatic except at the shock front, including those resonances which appear momentarily straddling the rarefied region behind the shock.  Even in the cases where the two resonance regions in the rarefaction zone overlap, our numerical treatment shows that the flavor transformation from the first (smaller radius) resonance is effectively nullified by the second.  If one assumes that the density profile behind the shock includes significant stochastic fluctuations, then this prescription is changed somewhat.  In the case where \lq\lq flavor depolarization\rq\rq\ is operative \cite{denfluct}, the resonances behind the shock would yield an equal mix of flavor states.  Such a treatment yields CC signals similar to that above, but with the late count rate reduced.  For the situation where the density fluctuations or gradients behind the shock are sufficiently strong to preclude adiabatic transformations entirely, the late time CC ratios are reduced even further, and the \lq\lq dips\rq\rq\ become steps.  (The signal from SN1987a has been analyzed with respect to the CPT-violating, inverted $\bar \nu$-mass scheme \cite{cpt&SN} without including shock effects, with a \lq\lq tension\rq\rq\ noted between the energy/count rate for the detected events and that expected from SN models.  If flavor depolarization or non-adiabatic level crossings occur behind the shock, this apparent discrepency may be reduced by the effects described here.)
 
This work was partially supported by NSF Grant PHY00-0099499 and the DOE SCiDAC Supernova Grant at UCSD, and University of California DOE contract W-7405-ENG-36.  We are indebted to A. Mezzacappa, G. C. McLaughlin, A. B. Balantekin, and W. C. Louis for insightful comments; and to J. R. Wilson and H. E. Dalhed for providing their recent supernova model output.

\end{document}